\documentclass[a4paper,11pt]{article}
\pdfoutput=1 % if your are submitting a pdflatex (i.e. if you have
\usepackage{jheppub} % for details on the use of the package, please
\usepackage[T1]{fontenc} % if needed

\newcommand{\n}[1]{\label{#1}}

\newcommand{\eq}[1]{(\ref{#1})}
\newcommand{\be}{\begin{eqnarray}}
\newcommand{\ee}{\end{eqnarray}}
\def\beq{\begin{eqnarray}}
\def\eeq{\end{eqnarray}}

\def\al{\alpha}
\def\be{\beta}

\def\ga{\gamma}
\def\de{\delta}

\def\ka{\kappa}

\def\om{\omega}
\def\ph{\varphi}

\def\Ga{\Gamma}

\title{%%%%   \boldmath
On Newtonian singularities in higher derivative gravity models}

\author[a]{Leonardo Modesto
}
\author[b,c]{Tib\'erio de Paula Netto,}
\author[c,d,e]{and Ilya L. Shapiro}

\affiliation[a]{Department of Physics \& Center for Field Theory and Particle Physics,
\\
Fudan University, 200433 Shanghai, China}
\affiliation[b]{Department of Physics, University of Alberta, Edmonton, Alberta, Canada T6G 2E1}
\affiliation[c]{Departamento de Fisica - ICE, Universidade Federal de Juiz de Fora,
\\
Campus da UFJF, CEP: 36036-900, Juiz de Fora, MG, Brazil}
\affiliation[d]{D\'epartement de Physique Th\'eorique and Center for
Astroparticle Physics,
\\
Universit\'e de Gen\`eve,
24 quai Ansermet, CH–1211 Gen\'eve 4, Switzerland}
\affiliation[e]{Tomsk State Pedagogical University and Tomsk State University}

\emailAdd{lmodesto@fudan.edu.cn}
\emailAdd{tiberiop@fisica.ufjf.br}
\emailAdd{shapiro@fisica.ufjf.br}

\abstract{We consider the problem of Newtonian singularity
in the wide class of higher derivative gravity models, including the
ones which are renormalizable and super-renormalizable at the quantum
level. The simplest version of the singularity-free theory has four
derivatives and is pretty well-known. We argue that in all cases of
local higher-derivative theories, when the poles of the propagator are
real and simple, the singularities disappear due to the cancelation
of contributions from scalar and tensor massive modes.}

\begin{document}
\maketitle
\flushbottom

\section{Introduction}
\label{sec:intro}

The singularities in the relevant solutions of general relativity
indicate the limits of applicability of the theory. For this reason
they represent an important motivation to study quantum effects, which
are supposed to make the theory free of singularities. For instance,
in cosmology we know that this really happens, since in the framework
of the complete, quantum theory-based, non-local version of the
Starobinsky model \cite{star,FHH} there is no initial singularity
\cite{Ander}. The exploration of the same issue in the black hole
case is much more difficult (see, e.g., \cite{FrolVilk} and also
\cite{FroSha}) and at present there is
no comprehensive investigation of the problem in this case. On the
other side, the singularity which looks very similar to the one in
the black hole solution can be met already in Newton gravity, in the
case of a single point-like particle. In the recent works (see also
\cite{Modesto,BMS,BGKM}), the Newtonian singularity problem has been
addressed in the framework of non-local gravity. The same non-local
model has been suggested earlier by Tomboulis \cite{Tomboulis}
as a version of super-renormalizable and ghost-free theory. It turns out
that at least some special version of the theory, with exponential
type of non-locality, the theory is really singularity-free. Let us
note that much earlier, the same result concerning the absence of
Newtonian singularity has been obtained in \cite{Tseytlin-95} for a
modified low-energy string effective action, which is essentially
equivalent to the model of \cite{Tomboulis} and \cite{Modesto,BMS,BGKM}.
In the present work we shall extend this result for a set of local
higher-derivative models suggested in \cite{highderi}.

In order to verify the presence of Newtonian singularity, one
can consider metric fluctuations around Minkowski space-time,
$g_{\mu\nu} =  \eta_{\mu\nu} + h_{\mu\nu}$. Then the linearized
Lagrangian provides the IR Newtonian limit in the amplitude
corresponding to the one-graviton exchange between two static
masses. Since the Newtonian gravity comes from the linear
approximation on the flat background, at the covariant
level the result is completely determined by the terms
of up to the second order in curvature tensor.

Consider the general version of higher derivative gravitational
action of up to the second order in curvature tensor, but without
restrictions on the number of derivatives. The
corresponding action has the form
%\beq
%\n{act}
%S = - \frac{1}{\ka} \int d^4 x \sqrt{-g} \,
%\Big\{ \, \frac{R}{2}
%- R \, \frac{F_1(\Box)}{4} \, R
%- R_{\mu\nu} \, \frac{F_2 (\Box) }{4} \, R^{\mu\nu}
%- R_{\mu\nu\al\be} \, \frac{F_3 (\Box) }{4} \, R^{\mu\nu\al\be}
%\Big\}\,,
%\eeq
\beq
\n{act}
S =  \frac{1}{4 \ka} \int d^4 x \sqrt{-g} \,
\Big\{ - 2  R
+ R \, F_1(\Box) \, R
+ R_{\mu\nu} \, F_2 (\Box)  \, R^{\mu\nu}
+ R_{\mu\nu\al\be} \, F_3 (\Box)  \, R^{\mu\nu\al\be}
\Big\}\,,
\eeq
where we used notation \ $\ka = 8 \pi G$.
Let us note that the cosmological constant is not included, because
it is known to be very small and also does not affect singularity in
the Newtonian potential of the point-like mass.

The expressions such as \eq{act} emerge naturally in different
physical situations. To start with, the one-loop semiclassical
corrections to the gravity action produce the form factors which
have exactly the form (\ref{act}), with the non-polynomial functions
$F_{1,2,3}=F_{1,2,3}(\Box/m^2)$, typically with the logarithmic
asymptotics in the far UV \cite{apco}. The constant values of
$F_{1,2,3}$ correspond to the well-known fourth-derivative models
of renormalizable quantum gravity \cite{Stelle-77}, with existing
extensive discussion of classical properties in the literature
starting from \cite{Stelle-78}. Furthermore, the polynomial form
of the same functions $F_{1,2,3}$, if being introduced
into the classical action, leads to super-renormalizable models of
quantum gravity \cite{highderi}, even if the ${\cal O}(R_{...}^3)$
and other higher-order (corresponding to the order of polynomials)
terms are included. In both these cases, however, the spectrum of the
theory has a set of massive spin-2 excitations, some of them are always
unphysical ghosts with negative kinetic energy. The problem of
ghosts has a long and interesting history, but since it is not the
subject of the present work, let us just readdress the reader to
the recent papers \cite{GW-HDQG} of one of us for a brief review and
further references. The models of \cite{highderi} were promoted to
the ghost-free non-polynomial form in \cite{Tomboulis}, but the
quantum properties of this version, such as (super)renormalizability
are not clear yet. One can see the discussion of this question in
\cite{GW-HDQG} and also parallel consideration in \cite{Modesto-QG}.

The functions $F_{1,2,3}$ used in \cite{Tomboulis} to avoid
ghosts are exponential, quite different from the asymptotically
logarithmic form factors of one-loop semiclassical terms of \cite{apco},
from the constants in the renormalizable gravity case \cite{Stelle-77},
and from the polynomial form of a super-renormalizable quantum gravity
models of \cite{highderi}.
As we have mentioned above, recently the exponential form of
these functions has been used in \cite{BGKM,BMS} to cure the
singularity of the Newtonian point-like solution. This result
looks quite remarkable. One may think that it indicates that the
strong in UV, exponential form-factors are very special and that
they are necessary to cure the singularity. If it is really so,
this would mean that there is some fundamental physical reason for
the absence of singularity in the strongly growing in UV functions
$F_{1,2,3}$. Then one may expect that the same will happens also
in the non-linear regime, when we deal with the full black-hole
solution, being it Schwarzschild or Kerr.

In the present work we are going to show that the singularity-free
solutions are possible not only for the exponential functions
$F_{1,2,3}$, but also for the polynomial versions of the
functions $F_{1,2,3}$, including the constant functions
(which is the very well-known result of \cite{Stelle-78}). Therefore,
in this work our purpose is to explore the singularity problem in a
different class of higher derivative theories compared to the ones
which were considered in  \cite{Stelle-78} and \cite{BGKM}. The paper
is organized as follows.
In Sect. 2 we present the general theoretical background of the
problem. Sect. 3 contains the main results, including the proof
of the non-singular behavior of local higher derivative gravity
theories with a real spectrum. In Sect. 4 we discuss the role of
tensor and scalar ghosts in the cancellation of Newtonian
singularity and, also, the existing relation between this
cancellation and (super)renormalizability of the corresponding
quantum theory. Finally, in Sect. 5 we draw our
conclusions and also present some discussions.

%%%%%%%%%%%%%%%%%%%%%%%%%%%%%%%%%%%%%%%%%%%%%%%%%%%%%%
%%%%%%%%%%%%%%%%%%%%%%%%%%%%%%%%%%%%%%%%%%%%%%%%%%%%%%
%%%%%%%%%%%%%%%%%%%%%%%%%%%%%%%%%%%%%%%%%%%%%%%%%%%%%%
%%%%%%%%%%%%%%%%%%%%%%%%%%%%%%%%%%%%%%%%%%%%%%%%%%%%%%
\section{Modified Newtonian limit}

The Newtonian limit means static weak-field approximation. So,
we consider metric fluctuations around Minkowski space-time
\beq
\n{glinear}
g_{\mu\nu} =  \eta_{\mu\nu} + h_{\mu\nu}\,.
\eeq
To find linearized field equations we need to consider only those terms
in the action which are of the second order in the perturbations
$h_{\mu \nu}$.
The following relevant observation is in order. By means of the
Bianchi identities and integrations by parts one can prove that
for any integer $N$
\beq
\n{gb}
\int d^4 x \sqrt{-g} \,\Big\{
R_{\mu\nu\al\be}\Box^N R^{\mu\nu\al\be}
-4R_{\mu\nu}\Box^N R^{\mu\nu}
+ R\Box^N R\Big\} \,=\,{\cal O}(R_{...}^3) \,=\,{\cal O}(h^3)\,.
\eeq
Assuming that the functions  $F_{1,2,3}$ admit an expansion into
power series in $\Box$, one comes to the conclusion that the
Riemann-squared term is not relevant in the linear regime.
Then one can simply trade $F_{1,2,3} \to {\tilde F}_{1,2,3}$,
where ${\tilde F}_1 = F_1-F_3$, ${\tilde F}_2 = F_2+4F_3$ and
${\tilde F}_3 = 0$. In what follows we effectively use the relations
for ${\tilde F}_{1,2,3}$, but do not write the tildes for
simplicity of notations. So, from now on, \ $F_3(\Box) \equiv 0\,$.

Performing the expansion in $h_{\mu\nu}$, the bilinear
part of the action \eq{act} is given by
\beq
&& \hspace{-2cm} \mathcal{L}_{\rm quadr} =
- \frac{1}{4 \kappa} [ h^{\mu \nu} \Box h_{\mu \nu} + A_{\nu}^2
+ (A_{\nu} - \phi_{, \nu})^2 ]  \nonumber \\
&&  \hspace{-0.3cm}
- \frac{1}{16 \kappa} \Big[
-
\Box   h_{\mu \nu}  F_2( \Box) \Box h^{\mu \nu}
+A^{\mu}_{, \mu}  F_2( \Box) A^{\nu}_{, \nu}
+ F^{\mu \nu}  F_2( \Box) F_{\mu \nu} \nonumber \\
&& \hspace{-0.3cm}
- (A^{\alpha}_{, \alpha} - \Box \phi) (F_2( \Box)
+  4 F_1( \Box) ) (A^{\beta}_{, \beta} - \Box \phi)
\Big]\,,
\label{quadratic2}
\eeq
where the vector and antisymmetric tensors are below defined
in terms of the gravitational fluctuation,
\beq
%&&
A^{\mu} = h^{\mu \nu}_{\,\,\,\, , \nu} \, , \,\,\,\, %\nonumber\\&&
\phi = h^{\mu}_{\mu} \,\,~~(\mbox{trace of} \,\,h_{\mu \nu}) \, , \,\,\,\, %\nonumber \\&&
F_{\mu \nu}= A_{\mu , \nu} - A_{\nu, \mu} \, .
\eeq

Inverting the quadratic operator in (\ref{quadratic2}),
we find the following two-point function in momentum space
in terms of spin-2 projector
$P^{(2)}$ and scalar projector $P^{(0-s)}$ (see, e.g.,
\cite{Stelle-77} or \cite{book} for details),
\beq
 \hspace{-0.3cm}
 {G}_{2}(k) =
  %\frac{\xi (2P^{(1)} +
%\bar{P}^{(0)} ) }{2 k^2 \, \omega( k^2/\Lambda^2)}
 \frac{P^{(2)}}{k^2 \left( 1 +  k^2  {F_2(-k^2)}/{2} \right) }
%\nonumber \\
%&&\hspace{-1cm}
- \frac{P^{(0 - s)}}{2 k^2 \left[ 1 -  k^2  \left(   F_2(-k^2) + 3 F_1(-k^2) \right) \right] } \, . \label{propagator}
\eeq

Since we are interested to find a solution for the point-like
static mass source, let us take
\beq
T_{\mu\nu} = \rho\, \de_\mu^0 \de_\nu^0
=
M \,\de^3({\bf r})\, \de_\mu^0 \,\de_\nu^0 \,.
\label{Tmn}
\eeq
One can then easily calculate the solution of the linear equations of motion coming from
(\ref{quadratic2}) by means
of the Fourier transform of the potential.
The Lagrangian for the graviton fluctuation and matter source reads
\beq
\mathcal{L}_h =
%\frac{1}{2}
h_{\mu \nu} (G_2^{-1} )^{\mu \nu, \rho \sigma} h_{\rho \sigma} -
%\frac{\sqrt{4 \kappa}}{2}
4\ka h_{\mu \nu} T^{\mu\nu}.
\eeq
Therefore in short notation,
\beq
\hat{h}= 2 \ka \, \hat{G_2} \, \hat{ T } \,\,\,\,
\Longrightarrow \,\,\,\, \varphi( r ) = - \frac{h_{00}}{2} \, .
\eeq
In this way, after
some algebra it is possible to express the potential as
\beq
\n{new-int}
&& \hspace{-1.2cm}
\ph (r)=
- \frac{2 G M}{\pi r}
\int_0^ \infty \frac{dp}{p} \sin ( p r )
\left\{ \frac{4}{3  \left( 1 +  p^2  {F_2}/{2} \right)} - \frac{1}{3 \left[ 1 -  p^2  \left(   F_2 + 3 F_1 \right) \right] }
\right\}  .
%\frac{(a-2c)\, \sin pr}{a(a-3c)}\,,
%\\
%&& \psi (r)
%=
%\frac{4 G M}{\pi r}
%\int_0^ \infty \frac{dp}{p}
%\frac{c \, \sin pr}{a(a-3c)}\,,
%\n{new-new-int}
\eeq
where $F_1 = F_1(-p^2)$ and $F_2 = F_2(-p^2)$.
We can introduce the short notation
\beq
\label{ac}
a(\Box) \equiv 1 -   \Box \, {F_2(\Box) }/{2} \,\,\, {\rm  and } \,\,\,
 c(\Box) \equiv 1 +  \Box \, \left(F_2(\Box) + 3 F_1(\Box) \right),
 \eeq
  for future reference.
For the special case where $a=c$ the potential is given by
\beq
\n{newton}
\ph (r) = %\psi (r) =
- \frac{2 \, G m}{\pi r}
\int_0^\infty \frac{dp}{p}
\,\frac{\sin (pr)}{a(-p^2)}
\,.
\eeq
Some extra notes about the $a=c$ case are in order. This condition
can be achieved if we choose the functions $F_i(\Box)$ according to
\beq
F_1(\Box) = \frac{a(\Box)-1}{\Box}\,,
\qquad
F_2(\Box)= - 2 F_1(\Box)\,,
\qquad
F_3(\Box) = 0\,.
\label{eq N3}
\eeq
This means, for the non-linear case, the special form of the
higher derivative part of the gravitational action
\beq
\n{maction}
S = -\, \frac{1}{2 \ka}
\int d^4 x \sqrt{-g}\,\,
\Big\{ R
+ \, G_{\mu\nu} \,\frac{ a(\Box) -1}{\Box}\, R^{\mu\nu}
\Big\}\,,
\label{eq N4}
\eeq
where $G_{\mu\nu}=R_{\mu\nu}-\frac{1}{2}g_{\mu\nu}R$
is the Einstein tensor. It is important to stress that, despite the
relation \eq{gb} holds in the linear approximation, in the non-linear
regime, $F_3(\Box) = 0$ can be achieved only in an {\it ad hoc} manner,
exactly as the second relation in \eq{eq N3}.
%%%%%%%%%%%%%%%%%%%
In what follows we will not apply the constraint  \ $a=c$ \ and will
mainly deal with the general case, except some places where it is
specially indicated.

The propagator of the gravitational field in the theory
\eq{maction} simplifies to the following form,
\beq
G_2(k)
=
\frac{1}{k^2 \,a(-k^2)}\,
\Big[ P^{(2)} - \frac{1}{2} P^{(0-s)} \Big]\,,
\label{eq N5}
\eeq
which has an algebraic structure that does not depend too much on
the form of the function $a(-k^2)$, since the last enters this
expression as an overall factor. The general relativity propagator
can be recovered if we set \ $a = 1$. Then, in order to have the
correct general relativity limit, one has to assume that in the
infrared, when $\,k^2\to 0$, this function must satisfy the
condition \ $a(-k^2)\to 1$. This requirement means that \
$a(\Box)$ \ should be a non-singular analytic function at \
$k^2=0$ and cannot contain non-local operators such as \ $1/\Box$.
Furthermore, if the residue of the $P^{(2)}$-term coefficient at
$k^2=0$ is negative, the theory contains a higher derivative ghost.
On the other hand, by choosing $a(\Box)$ to be an entire function,
one can construct a theory being free from higher derivative
ghosts \cite{Tomboulis}.

Essentially the same example of entire function has been considered
in Ref. \cite{Modesto, BGKM, ModestoHD} (see also \cite{Nicolini:2005vd,Modesto:2010uh}, including for the non-commutative
geometry case), namely
\beq
a (\Box) = e^{-\Box/m^2}\,.
\n{amazum}
\eeq
For the function \eq{amazum} the solution for the modified
Newtonian potential \eq{newton} is
\beq
\ph (r) = - \frac{GM}{r} \, \mbox{erf}
\left( \frac{m r }{2} \right)\,.
\n{eq N6}
\eeq
Since $\mbox{erf} \, (r)\to r$ when $r \to 0$,
the modified Newtonian potential has a non-singular behavior.

Is it true that the singularity disappears due to the non-locality
and that the effect depends on the presence of the exponential form
factor? In order to answer this question, in the next section we are
going to construct other examples of the functions $F_{1,2,3}(\Box)$
leading to the non-singular at \ $r=0$ \ modified Newtonian limit
and also have a correct infrared behavior at \ $r \to \infty$.

%%%%%%%%%%%%%%%%%%%%%%%%%%%%%%%%%%%%%%%%%%%%%%%%%%%
%%%%%%%%%%%%%%%%%%%%%%%%%%%%%%%%%%%%%%%%%%%%%%%%%%%
%%%%%%%%%%%%%%%%%%%%%%%%%%%%%%%%%%%%%%%%%%%%%%%%%%%
\section{Polynomial functions}

As we have already mentioned in the Introduction, the main
advantage of the polynomial form factors is that the corresponding
theory is (super)renormalizable, that is not certain yet for the
nonlocal ghost-free models. Consider the most general local
action \cite{highderi}
\beq
&& \hspace{-0.9cm}
S = \frac{1}{4\ka} \int d^4 x \sqrt{-g} \,\, \Big\{
-2 R + \al_0 R_{\mu \nu}^2 + \be_0 R^2
+ \ga_0 R_{\mu\nu\al\be}^2 + \dots
\nonumber
\\
&& + \al_1 R_{\mu \nu} \Box R^{\mu \nu} + \be_1 R \Box R
+\ga_1 R_{\mu \nu \al \be} \Box R^{\mu \nu \al \be}
\,+\, {\cal O} (R^3_{...})
+ \dots
\n{local-act}
\\
&& + \al_N R_{\mu \nu} \Box^N R^{\mu \nu} + \be_N R \Box^N R
+ \ga_N R_{\mu \nu \al \be} \Box^N R^{\mu \nu \al \be}
+ \dots + {\cal O} (R^{N+2}_{...})
\Big\}
\,.
\nonumber
\eeq
As we have already explained above, only the terms quadratic in
curvature tensor may be relevant for deriving the modified Newtonian
potential. Moreover, due to the relation (\ref{gb}) one can safely
omit the terms with the squares of the Riemann tensor, so the result
will depend only on the coefficients $\,\al_i\,$ and $\,\be_i$, with
$\,i=0,1,...,N$.

The $N = 0$ model is the 4th-order gravity \cite{Stelle-77}.
For this case the solution for the modified Newtonian potential is
pretty well-known \cite{Stelle-77,Stelle-78},
\beq
\n{eq9}
\ph (r) = - GM \left(\frac{1}{r} - \frac{4}{3} \frac{e^{-m_{(2)} r}}{r}
+ \frac{1}{3} \frac{e^{-m_{(0)} r}}{r} \right)
\,.
\eeq
The mass parameters are defined by $m_{(2)} = (\frac{1}{2}\al_0)^{-1/2}$
for the tensor mode and by $m_{(0)} = [-(3\be_0+\al_0)]^{-1/2}$ for the
scalar mode. The scalar sector has a gauge-fixing ambiguity \cite{book},
but this has no importance for the scattering amplitude behind the
result \eq{eq9}. For the sake of simplicity we can say that the two
parameters $m_{(2)}$ and $m_{(0)}$ correspond, respectively, to the
masses of tensor and scalar massive degrees of freedom in the
propagator of the theory.

At large distances the effects of the Yukawa corrections in
\eq{eq9} disappear and one meets a standard Newton limit in
the gravitational potential. On the other hand, in the short-distance
regime the situations depends on the coefficients $\al_0$ and $\be_0$.
At the origin $r = 0$, expanding the exponential into power series
one can easily check that the contributions of higher derivative terms
to the Newtonian potential make it regular. The modified potential
tends to the constant value
\beq
\ph(r) =
-\,\frac{1}{3}\,G M\,\big[4 m_{(2)}\,-\, m_{(0)}\big]
\,+\,{\cal O}(r)\,.
\label{const}
\eeq
This well-known example shows that the theory without
any kind of non-locality can be free from the Newton singularity. The
singularity cancellation occurs because the zero-order terms of the
two different Yukawa potentials combine exactly into the coefficient
$-4/3 + 1/3=-1$, to cancel the original Newtonian term.

The solution for the Newtonian potential in the theory which has only
Einstein-Hilbert and the square of scalar curvature terms in the action
(i.e., when $\al_i = \ga_i = 0$ for $i = 0,1,\cdots, N$ in \eq{local-act})
was previously considered in Ref. \cite{Schmidt}. It was shown that
in this case the modified Newtonian potential gains a higher derivative
contribution given by a sum of Yukawa potentials, namely
\beq
\ph (r) = - \, GM \left(\frac{1}{r}
+ \sum_{i=0}^{N} \, \frac{c_i}{r}\,e^{-m_{(0)i} \, r} \right)
\,,
\eeq
where the coefficients \ $c_i$ \ satisfy the condition
\beq
\n{constraint}
\sum_{i=0}^{N} c_i = \frac{1}{3}\,.
\eeq

For the full theory with at least $\al_N \neq 0$ we expect a
number of new terms with Yukawa potentials coming for the
Ricci tensor-squared terms.
And if the coefficients of these Yukawa potentials satisfy some
kind of relation like \eq{constraint} but with the sum of
coefficients equal to $-4/3$, then the Newtonian potential
is singularity free for the general local higher derivative
gravitational action of the form (\ref{act}). The proof of this
statement is the main purpose of this section.

In the linear regime the action \eq{act} is equivalent to
the action \eq{local-act} with
\beq
&& \n{loc-f2}
F_2 (\Box) = \al_0 + \al_1 \Box
+ \cdots + \al_N \Box^N
\,,
\\
&& \n{loc-f1}
F_1 (\Box) = \be_0 + \be_1 \Box
+ \cdots + \be_N \Box^N
\,.
\eeq
Consider the integral \eq{new-int} for the Newtonian potential
$\ph(r)$. It is easy to see that for the case of our interest
there is the following relation (\ref{new-int}), (\ref{ac}):
\beq
\n{iden}
%\frac{1}{2}
 \left( \frac{4}{3 a} - \frac{1}{3c} \right) =
%\frac{a-2c}{a(a-3c)} = \frac{1}{2}
\left[ \frac{4}{3} \frac{1}{P_{2N+2}}
- \frac{1}{3} \frac{1}{Q_{2N+2}} \right]\,,
\eeq
where $P_{2N+2}$ and $Q_{2N+2}$ are polynomials of $\Box$ of
the corresponding order. Due to Eqs. \eq{loc-f2}$-$\eq{loc-f1},
the functions $P_{2N+2}$ and $Q_{2N+2}$ can be written in terms
of the coefficients $\al_N$ and $\be_N$ as
\beq
\n{pp}
P_{2N+2} = 1 + \frac{1}{2}\, \big[\al_0 p^2
- \al_1 p^4 + \cdots + (-1)^N \, \al_N \, p^{2N+2}\big]\,,
\eeq
\beq
\n{pq}
Q_{2N+2} = 1 - ( 3\be_0 + \al_0) p^2
+ ( 3\be_1 + \al_1)  p^4 + \cdots + (-1)^{N+1}
\, ( 3\be_N + \al_N) \,  p^{2N+2}\,.
\eeq
Let us assume that the coefficients of the polynomials \eq{pp},
\eq{pq} do not vanish, i.e, $\al_i \neq 0$ and $\al_i + 3 \be_i
\neq 0$ for $i = 0,1,\cdots,N$. According to the fundamental
theorem of algebra, the polynomials $P_{2N+2}$ and $Q_{2N+2}$
can be factorized as
\beq
\n{for1}
P_{2N+2} = \frac{1}{m_{(2)0}^2 m_{(2)1}^2 \cdots m_{(2)N}^2}
\times (p^2 + m_{(2)0}^2)\times (p^2 + m_{(2)1}^2) \times
\cdots \times (p^2 + m_{(2)N}^2)\,,
\eeq
\beq
\n{for2}
Q_{2N+2} = \frac{1}{m_{(0)0}^2 m_{(0)1}^2 \cdots m_{(0)N}^2}
\times (p^2 + m_{(0)0}^2)\times (p^2 + m_{(0)1}^2) \times
\cdots \times (p^2 + m_{(0)N}^2)\,.
\eeq
Here the square of the roots of Eqs. $P_{2N+2} = 0$ and $Q_{2N+2} = 0$
are $\,-m_{(2)N}^2$ and $\,-m_{(0)N}^2$, correspondingly.

In what follows we assume that
by adjusting the coefficients $\al_N$ and $\al_N+3\be_N$ of
subleading terms of the polynomials $P_{2N+2}$ and $Q_{2N+2}$
it is possible to provide that all $m_{(k)j}$
are real quantities and
\beq
&&
\n{massrel}
0< m_{(k)0}^2 < m_{(k)1}^2 < \cdots < m_{(k)N}^2\,,
\\
&&
m_{(k)i} \neq m_{(k)j} \,,
\qquad i \neq j
\eeq
for $k = 0,2$. The last condition means that all
the poles of the propagator are simple.
From the physical side $m_{(2)N}$ and $m_{(0)N}$ corresponds to
the masses of spin-2 and spin-0 massive extra degrees of freedom
in the propagator of the theory \eq{local-act}.

Using the identity \eq{iden} and formulas \eq{for1}, \eq{for2}
the Newtonian potential \eq{new-int} can be cast into the form
\beq
\n{phI1I2}
\ph (r) = - \frac{2GM}{\pi r} \left[ \frac{4}{3} \, I_{(2)}
- \frac{1}{3} \, I_{(0)} \right],
\eeq
where
\beq
I_{(2)} = \int_0^\infty dp
\, \, \frac{(m_{(2)0}^2 m_{(2)1}^2 \cdots m_{(2)N}^2 )\sin (pr)}
{p(p^2 + m_{(2)0}^2)(p^2 + m_{(2)1}^2)
\cdots (p^2 + m_{(2)N}^2)}
\eeq
and
\beq
I_{(0)} = \int_0^\infty dp
\, \, \frac{(m_{(0)0}^2 m_{(0)1}^2 \cdots m_{(0)N}^2 )\sin (pr)}
{p(p^2 + m_{(0)0}^2)(p^2 + m_{(0)1}^2)
\cdots (p^2 + m_{(0)N}^2)}.
\eeq

To evaluate the integrals $I_{(2)}$, $I_{(0)}$ we perform an analytic
continuation $p\rightarrow z$ to the complex plane ${\mathbb C}$.
Then the integral $I_{(2)}$ can be written as
\beq
\n{I2W}
I_{(2)} = \frac{W_1 - W_2}{4i},
\eeq
where
\beq
W_1 = \oint_\Ga dz
\, \, \frac{(m_{(2)0}^2 m_{(2)1}^2 \cdots m_{(2)N}^2 ) \, e^{i z r}}
{z(z^2 + m_{(2)0}^2)(z^2 + m_{(2)1}^2)
\cdots (z^2 + m_{(2)N}^2)},
\eeq
\beq
W_2 = \oint_\Ga dz
\, \, \frac{(m_{(2)0}^2 m_{(2)1}^2 \cdots m_{(2)N}^2 ) \, e^{-i z r}}
{z(z^2 + m_{(2)0}^2)(z^2 + m_{(2)1}^2)
\cdots (z^2 + m_{(2)N}^2)}.
\eeq
Since the masses $m_{(2)j}$ are different, the integrals $W_1$, $W_2$
have simple poles at the points $z=0$ and $z^2 = - m_j^2$, where
$j = 0,1,\cdots,N$. \ Let $\Ga$
be a positively oriented simple closed path in ${\mathbb C}$ which
passes on the left of the poles on the lower half plane
$z = -i m_{(2)j}$ and on the right of the poles at the points
$z = 0$ on the upper half plane $z = + i m_{(2)j}$.

%% \begin{figure}[tbp]
%% \centering % \begin{center}/\end{center} takes some additional vertical space
%% \includegraphics[width=.45\textwidth,trim=0 380 0 200,clip]{img1.pdf}
%% \hfill
%% \includegraphics[width=.45\textwidth,origin=c,angle=180]{img2.pdf}
%% "\includegraphics" is very powerful; the graphicx package is already loaded
%% \caption{\label{fig:i} Always give a caption.}
%% \end{figure}

%%%%%%%%%%%%%%%%%%%%%%%%%%%%%%%%%%%%%%%%%%%%%%%%%%%%%%%%%%%%
\begin{figure}[tb]
\vskip -0.5cm
\centerline{
\resizebox*{4.6 cm}{!}{\includegraphics{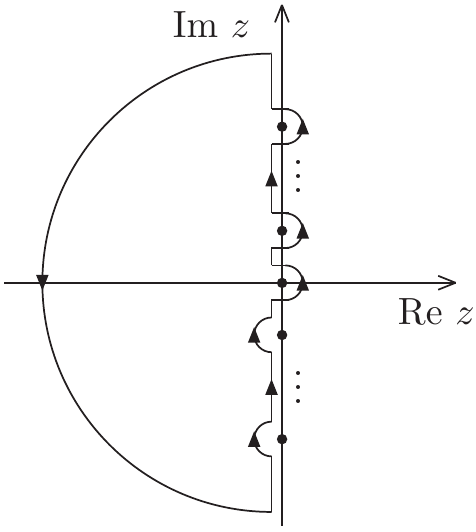}}
\resizebox*{4.2 cm}{!}{\includegraphics{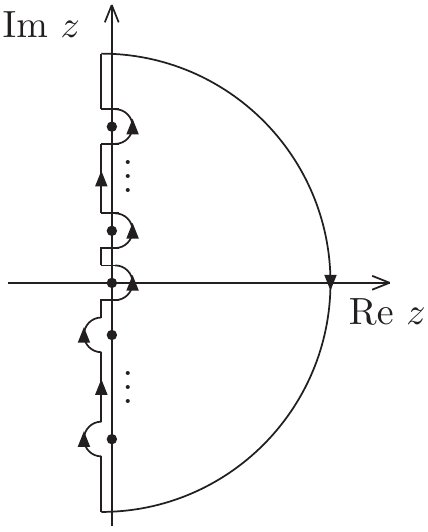}}}
\begin{quotation}
\caption{The first and second curves of the integration on the
complex plane. On the left the poles at $z=0$ and $z=+im_{(2)N}$ are
inside the contour. On the right the poles at $z=-im_{(2)N}$ are
inside the contour.}
\end{quotation}
\end{figure}
%%%  Figure 1.1
%%%%%%%%%%%%%%%%%%%%%%%%%%%%%%%%%%%%%%%%%%%%%%%%%%%%%%%%%%%%%%%%%

For $W_1$ which contains $e^{irp}$, the contour $\Ga$ should be
chosen in such a way that it encircles the poles at $z=0$ and
$z = + i m_{(2)j}$.
One can see the left plot of Fig. 1 for the illustration.
Then, using the Cauchy's residue theorem we find
\beq
&& \n{w1}
\hspace{-0.7cm}
W_1 = + \,\, 2 \pi i
\Bigg\{ \mbox{Res} \Big[
\frac{(m_{(2)0}^2 \cdots m_{(2)N}^2) \, e^{i z r}}
{(z^2 + m_{(2)0}^2)\cdots (z^2 + m_{(2)N}^2)}
\, , \, z = 0 \Big]
%\nonumber
\\
&& \hspace{0.45cm}
+ \mbox{Res} \Big[
\frac{(m_{(2)0}^2 \cdots m_{(2)N}^2) \, e^{i z r}}
{z(z + i m_{(2)0})\cdots (z^2 + m_{(2)N}^2)}
\, , \, z = + i m_{(2)0} \Big]
\nonumber
\\
&& \hspace{0.45cm}
+ \cdots +  \mbox{Res} \Big[
\frac{(m_{(2)0}^2 \cdots m_{(2)N}^2) \, e^{i z r}}
{z(z^2 + m_{(2)0}^2)\cdots (z  + i m_{(2)N})}
\, , \, z = + i m_{(2)N} \Big] \Bigg\}
\nonumber
\\&& \hspace{-0.7cm}
= \!
1 \! - \frac{(m_{(2)1}^2 \! \cdots m_{(2)N}^2) \,  e^{-m_{(2)0}r}}
{2(m_{(2)1}^2-m_{(2)0}^2)\cdots (m_{(2)N}^2-m_{(2)0}^2) }
+ \! \cdots \!
-
\frac{(m_{(2)0}^2 \! \cdots m_{(2)N-1}^2) \, e^{-m_{(2)N}r}}
{2(m_{(2)0}^2-m_{(2)N}^2) \cdots (m_{(2)N}^2 - m_{(2)N-1}^2) }\,.
\nonumber
\eeq
The integral $W_1$ is calculated in counterclockwise direction
which we define to be positive.

For $W_2$ which has $e^{-irp}$, the path $\Ga$ is chosen in
such way that encircles the poles at $z = - i m_{(2)j}$.
One can see the right plot of Fig. 1 for the illustration.
The integral is evaluated in clockwise direction, then we find
\beq
&&
\n{w2}
\hspace{-0.5cm}
W_2 = - \,\, 2 \pi i \Bigg\{
+ \mbox{Res} \left[
\frac{(m_{(2)0}^2 \cdots m_{(2)N}^2) \, e^{-i z r}}
{z(z - i m_{(2)0})\cdots (z^2 + m_{(2)N}^2)}
\, , \, z = - i m_{(2)0} \right]
%\nonumber
\\
&&\hspace{-0.5cm}
 + \cdots +  \mbox{Res} \left[
\frac{(m_{(2)0}^2 \cdots m_{(2)N}^2) \, e^{-i z r}}
{z(z^2 + m_{(2)0}^2)\cdots (z  - i m_{(2)N})}
\, , \, z = - i m_{(2)N} \right] \Bigg\}
\nonumber
\\
&& \hspace{-0.5cm}
=
\frac{(m_{(2)1}^2 \cdots m_{(2)N}^2) \,\, e^{-m_{(2)0}r}}
{2(m_{(2)1}^2-m_{(2)0}^2)\cdots (m_{(2)N}^2-m_{(2)0}^2) }
+ \cdots
%\nonumber\\&&
+
\frac{(m_{(2)0}^2 \cdots m_{(2)N-1}^2) \,\, e^{-m_{(2)N}r}}
{2(m_{(2)0}^2-m_{(2)N}^2)\cdots (m_{(2)N}^2-m_{(2)N-1}^2) }
\,. \nonumber
\eeq

Now, by using Eqs. \eq{w1}, \eq{w2} and \eq{I2W} we obtain
\beq
\n{I2}
I_{(2)} = \frac{\pi}{2} \left[
1 - \sum_{i=0}^N \Big(
\prod_{j \neq i} \frac{m_{(2)j}^2}{m_{(2)j}^2-m_{(2)i}^2}
 \,e^{- m_{(2)i} r} \Big) \right].
\eeq
By an analogous consideration it is possible to evaluate $I_{(0)}$.
We are going to leave this calculation to be an exercise for an
interested reader, the answer is
\beq
\n{I0}
I_{(0)} = \frac{\pi}{2} \left[
1 - \sum_{i=0}^N \Big(
\prod_{j \neq i} \frac{m_{(0)j}^2}{m_{(0)j}^2-m_{(0)i}^2}
 \,e^{- m_{(0)i} r} \Big) \right].
\eeq

Finally, from \eq{phI1I2},\eq{I2} and \eq{I0} we arrive at the final
answer for the modified Newtonian potential
\beq
&&\hspace{-1.0cm}
 \n{pot}
\ph(r)
=
- \,GM \,\left\{ \frac{1}{r}
-\frac{4}{3} \sum_{i=0}^N \prod_{j \neq i}
\frac{m_{(2)j}^2}{m_{(2)j}^2-m_{(2)i}^2}
\frac{e^{- m_{(2)i} r}}{r} \right.
 \nonumber\\
&& \hspace{0.4cm}
\left. +
\frac{1}{3} \sum_{i=0}^N \prod_{j \neq i}
\frac{m_{(0)j}^2}{m_{(0)j}^2-m_{(0)i}^2}
\frac{e^{- m_{(0)i} r}}{r}   \right\}\,.
\eeq

Now let us study the behavior of potential \eq{pot}
near the origin. When $r \to 0$
\beq
\n{limit}
\ph (r) \to \frac{1}{r}
-\frac{4}{3r} \sum_{i=0}^N \prod_{j \neq i}
\frac{m_{(2)j}^2}{m_{(2)j}^2-m_{(2)i}^2}
+\frac{1}{3r} \sum_{i=0}^N \prod_{j \neq i}
\frac{m_{(0)j}^2}{m_{(0)j}^2-m_{(0)i}^2}
+ {\rm const}.
\eeq
For any set of numbers $a_j$ the following relation is valid:
\beq
\n{rel}
\sum_{i=0}^N \prod_{j \neq i}
\frac{a_j}{a_j - a_i} = 1\,.
\eeq
With this relation, one can see that the limit \eq{limit} goes
to a constant and the modified Newtonian potential is regular
at $r=0$.

To illustrate the consideration of this section, let us present
an exact solution for the sixth-order gravity, corresponding to
$N=1$. In this case the masses of the spin-2 particles are given by
\beq
 m_{(2)0}^2 = \frac{-\al_0
- \sqrt{\al_0^2+8\al_1}}{2 \al_1} \,, \,\,\,\,\,\,
%\eeq
%\beq
m_{(2)1}^2 = \frac{-\al_0
+ \sqrt{\al_0^2+8\al_1}}{2 \al_1}\,.
\eeq
In order for these solutions to define two different non-zero
real masses, the parameters should satisfy the conditions
\beq
\al_0 > 0\,, \qquad
\al_1 <0\,, \qquad
\al_0^2+8 \al_1 > 0 \,.
\eeq
For the massive scalar particle, defining
$\om_N \equiv 3 \be_N + \al_N$, we have
\beq
m_{(0)0}^2 = \frac{-\om_0 + \sqrt{\om_0^2 - 4 \om_1}}
{2 \om_1}
\,, \,\,\,\,\,
%\eeq
 %\beq
m_{(0)1}^2 = \frac{-\om_0 - \sqrt{\om_0^2 - 4 \om_1}}
{2 \om_1}
\,.
\eeq
For real different masses we need to impose
\beq
\om_0 < 0\,, \quad
\om_1 >0\,, \quad
\om_0^2-4 \om_1 > 0 \,.
\eeq
Since $\al_0$ must be positive and $\al_1$
must be negative, these relations are true only if
\beq
\be_0 < 0\,, \quad
\be_1 >0
\eeq
and if their absolute values satisfy
\beq
|\be_0| > \frac{1}{3}|\al_0|, \quad
|\be_1| > \frac{1}{3}|\al_1|.
\eeq

%%%%%%%%%%%%%%%%%%%%%%%%%%%%%%%%%%%%%%%%%%%%%%%%%%%%%%%%
%%%%%%%%%%%%%%%%%%%%%%%%%%%%%%%%%%%%%%%%%%%%%%%%%%%%%%%%
%%%%%%%%%%%%%%%%%%%%%%%%%%%%%%%%%%%%%%%%%%%%%%%%%%%%%%%%
\section{Ghosts and repulsion forces}

As we know from \cite{Stelle-77}, the cancellation of Newtonian
singularity in the four-derivative case is due to the opposite
signs of the contribution of graviton and scalar degree of freedom
from one side, and the massive tensor ghost from another side. It
would be interesting and useful to understand whether  a similar
relation takes place for the higher derivative models of
\cite{highderi}, especially in view of the absence of
singularities that takes place for the ghost-free theory of
\cite{Tseytlin-95} and \cite{Modesto, ModestoHD, Tomboulis}.

In the case of fourth-order gravity, the Eq. (\ref{eq9}) shows that
the massive spin-$2$ ghost particle contributes with an opposite sign,
 different from the contribution of graviton and scalar massive particle.
When a test particle is approaching to the origin $r = 0$ the gravitational
force applied to it tends to zero because the repulsive force due to the
ghosts cancels the attractive force of graviton plus an extra massive
scalar degree of freedom. Let us show that the same situation holds for
the more complicated case of superrenormalizable gravity theory. In this case,
again, one can say that all ghost particles contribute with repulsive
force, while the non-ghost degrees of freedom always contribute to the
attractive force.

To prove this statement, let us begin considering the scalar sector of the theory.
Consider the propagator of the scalar part,
\beq
G^{(0)}_2 (k) =
\left[\frac{A_0}{k^2 + m_{(0)0}^2} +
\frac{A_1}{k^2 + m_{(0)1}^2} + \cdots +
\frac{A_N}{k^2 + m_{(0)N}^2} \right] P^{(0-s)}
\,.
\eeq
according to \cite{highderi}, the residues of the propagator
satisfy $A_j \,.\, A_{j+1} <0$. For the scalar degree of freedom
we have $A_0>0$.  As a consequence, the residue $A_k$ with an
odd $k$ always has a negative sign and represents a ghost
particle, while the even components are always a non-ghost
degrees of freedom. It proves useful to rewrite the contribution
to the  gravitational potential coming from the $i$-th massive
scalar particle, that is the last term in Eq. (\ref{limit}), in
the form with an explicit sign dependence,
\beq
\n{posc}
\ph_{(0)i}(r)
=
-\frac{GM}{3} (-1)^{i} \prod_{j \neq i}
\left| \frac{m_{(0)j}^2}{m_{(0)i}^2}-1\right|^{-1}\,
\frac{e^{- m_{(0)i} r}}{r}\,.
\eeq
In the last equation we used the relation \eq{massrel} and the
sign of each product is shown explicitly. For and odd $i$ we
have a ghost, that is the sign in \eq{posc} is positive and
there is a repulsive potential. At the same time, the massive
healthy particles contribute to an attractive force.

For the tensorial part the situation is similar. The total
propagator of the spin-$2$ massive particles can be written as
\beq
G^{(2)}_2 (k) =
\left[\frac{B_0}{k^2 + m_{(2)0}^2} +
\frac{B_1}{k^2 + m_{(2)1}^2} + \cdots +
\frac{B_N}{k^2 + m_{(2)N}^2} \right] P^{(2)}\,,
\label{tens}
\eeq
where the residues satisfy $B_j \, .\, B_{j+1} <0$ \cite{highderi}.
Since for the spin-$2$ massive particles we have $B_0 < 0$, each
$B_k$ with an even index have negative sign and represent a ghost.
The gravitational potential for the $i$-th spin-$2$ massive particle
can be written as
\beq
\n{poten}
\ph_{(2)i}(r)
=
+\,\frac{4GM}{3} (-1)^{i} \prod_{j \neq i}
\left| \frac{m_{(2)j}^2}{m_{(2)i}^2}-1 \right|^{-1}\,
\frac{e^{- m_{(2)i} r}}{r}\,.
\eeq
For the ghost potentials, when $i$ is even, we have a positive sign
in \eq{poten} and, consequently, a repulsive force.

With the simple consideration presented above, we have shown that
for a point-like source the ghosts always induce a repulsive
Newtonian potential. As in the fourth-order gravity, in the
superrenormalizible models of \cite{highderi} the singularity of
the potential disappears because the repulsive force acting on a
test particle due to the ghosts cancels with the attractive force
of graviton and non-ghosts massive particles near $r=0$.

The main point of the above consideration is that the singularity
cancellation only occurs because for each massive spin-$2$ ghost
particle we have a non-ghost massive scalar, and vice-versa. This
structure of cancellation has an important consequence. If we
recast the relevant part of the action (\ref{act}) in the form
\beq
\n{act-W}
S = \frac{1}{4 \ka} \int d^4 x \sqrt{-g} \,
\Big\{- 2 R
+ R \, \Phi_1 (\Box) \, R
+ \, \frac{1}{2} \,\, C_{\mu\nu\al\be} \,\Phi_2 (\Box) \,C^{\mu\nu\al\be}
\Big\}\,,
\eeq
where $\Phi_2 =  F_2$, $\Phi_1=F_1+F_2/3$ and $C^{\mu}_{\,.\,\nu\al\be}$ is the Weyl tensor, then
where the form factor $\Phi_2$ alone will define the tensor sector
and the form factor $\Phi_1$ alone, the scalar sector. Imagine that
the two functions $\Phi_1$ and $\Phi_2$ are polynomials of the
{\it different}  orders. Then the pairs scalar particle - tensor
ghost and tensor particle - scalar ghost will be broken and there
will be no singularities cancellation. The effect takes place
{\it only} when the two polynomials are of the same order. It is
interesting that this corresponds, in principle, to the condition
of superrenormalizability as it follows from the consideration of
\cite{highderi}. In case of the different orders of the two
polynomials one meets non-homogeneous propagators and vertices
and it is certainly possible to have some diagrams with the
growing power counting index. This means that there is a direct
relation between the cancellation of Newtonian singularity and
quantum renormalizability properties. Of course, this relation is
a kind of a {\it post factum} feature, which may not have deep
physical meaning, but it gives, anyway, a certain hint to the
quantum properties of the ghost-free theory with an exponential
form-factor, as suggested in \cite{BGKM, ModestoHD}. As we have
noted in \cite{GW-HDQG}, the power counting in this theory is
indefinite, of the $\infty-\infty$ type, but we can not exclude
at the moment a consistent way to define a quantum field theory
of gravity with asymptotically exponential growth. On the other
hand, the theory is perfectly well defined for exponential form
factors asymptotically polynomial \cite{Modesto, Tomboulis}. In
this case the theory is unitary and superrenormalizable or finite
at quantum level \cite{Modesto, Tomboulis, ModestoLeslaw}.
At the same time, the absence of Newtonian singularity for the
class of theories in \cite{BGKM, ModestoHD} tells us that the UV
behavior of this theory is the right one, corresponding to the
(super)renormalizable models of quantum gravity. In our opinion,
this gives a strong hope and motivation to study the quantum UV
divergences of this model in more details than it was done until
now.

Another possibility to interpret the role of ghosts in the
cancellation of singularities concerns the proposal of
\cite{Hawking-H} that the consistent quantum theory must
describe ghosts not as individual particles, but as part of
a pair of ghost and graviton. As we have seen, this idea is
not working for the cancellation of Newtonian singularity and
one can easily show that it is nor working also for the
super-renormalizable models suggested in \cite{highderi}.
The consideration presented above shows that the role of the
ghosts and normal particles in the singularity
cancellation requires that these particles should actually enter
by the pairs of scalar particle plus tensor ghost and tensor
particle plus scalar ghost. This may mean that the proposal of
\cite{Hawking-H} should be modified accordingly.

%%%%%%%%%%%%%%%%%%%%%%%%%%%%%%%%%%%%%%%%%%%%%%%%%%%%%%%%
%%%%%%%%%%%%%%%%%%%%%%%%%%%%%%%%%%%%%%%%%%%%%%%%%%%%%%%%
%%%%%%%%%%%%%%%%%%%%%%%%%%%%%%%%%%%%%%%%%%%%%%%%%%%%%%%%
\section{Conclusions and discussions}

We have considered the problem of point-mass singularity in
the wide class of higher derivative models, including fourth
derivative ones, and the higher than four-derivative theories,
of the polynomial (superrenormalizable at quantum level) type.
The singularity in the modified Newtonian potential disappears
in the theories of (\ref{local-act}), due to the cancellation
of the contributions from scalar and tensor modes to the
Yukawa-type potential with the initial Newtonian singularity.

The cancellation can be easily provided in the ghost-free model
of \cite{Tomboulis,BGKM}, Eq. (\ref{amazum}), in the fourth
derivative model of \cite{Stelle-77} and in the superrenormalizible
gravity. For the last two cases the tensors and scalars contribute
respectively with coefficients $-4/3$ and $1/3$ near the origin,
and this leads to the cancellation of Newtonian singularity.
Compared to the case of exponential form factors considered in
\cite{BGKM}, we see that the presence of non-polynomial and
therefore non-local terms is not really necessary for the
cancellation of Newtonian singularity. At the same time, it is
remarkable that the local and non-local theories manifest the
same property. In particular, one may expect that the
non-singular feature of local theories will hold under
semiclassical \cite{apco} and quantum gravity corrections
to the terms quadratic in curvature. At the same time, the
definite answer to this question can be obtained only after
more detailed analysis, which we postpone for the future work.

The effect of singularity cancellation is essentially a linear
effect involving the independent contributions of scalars and
tensors. Therefore, it is not certain that the cancellation may
hold in this theories at the non-linear level, e.g., for the black
hole solutions. Of course, the singularity avoidance in black holes
by means of higher derivatives is natural and is expected to be
possible, but in order to verify this phenomenon one has to go
beyond the linear approximation, which works so well in the
modified Newtonian case.
%%%%%%%%%%%%%%%%%%%%%%%%%
Let us note that some aspects of the full non-linear solutions for
the static spherically symmetric case in the fourth-order gravity
($F_i =$ constant, in our notations) were considered in
\cite{Stelle-78}, then in \cite{FroSha} and, most recently, in
\cite{Stelle-2015}). Through the study of the asymptotic behavior
of the solutions it was shown that there are different families of
solutions, some equivalent to the Schwarzschild solution in GR.
According to \cite{Stelle-78}, one of these solutions is regular
at the origin. However, in order to determine whether the 
solutions that match to the modified Newton solution at infinity 
still have singularity at $r=0$ or not, a more detailed analytical 
or numerical investigation is needed.
%%%%%%%%%%%%%%%%%%%%%%%%%
In the more complicated theories with higher than four derivatives,
which we consider here, one can expect
a similar general situation in the non-linear regime and hence a
complicated analysis of the general field equations is necessary
to achieve concrete results about the black hole singularities.
The first step in this direction was done in \cite{Holdom-2002},
where the theory under investigation is apparently the one of
\cite{highderi}. Indeed, the comprehensive investigation of the
problem of $r=0$ singularity is very difficult and still not
completed task in both cases. An important aspect is that the
divergences of quantum theory are apparently related to the
Newtonian singularities and not to the ones at the non-linear
level of modified .

One may think about some relation between the absence of Newtonian
singularity in classical theory and asymptotic freedom at quantum
level. Such a relation would be somehow natural, because Newtonian
singularity is
indeed the simplest UV divergence due to the interaction. So,
when the singularity disappears, it looks like a kind of an
UV screening of the interaction \cite{cosimo}. However, our results show that
the relation with asymptotic freedom is not so relevant. Indeed,
the cancellation of singularity occurs in all superrenormalizable
models of \cite{highderi} if the massive spin-2 and spin-0 excitations
correspond to real simple poles. At the same time, most of the
coupling constants in the theories described in \cite{highderi}
are not renormalized. For example, in case of $N \geq 3$ the
one-loop $\be$-functions are exact and they are non-zero only
for the zero-, two- and four-derivative terms. Furthermore,
these $\be$-functions depend on the coupling in the highest
derivative sectors (but not on the gauge fixing!), and their
sign can be deliberately changed by tuning these highest
derivative couplings,
while the couplings in the higher than
four-derivative sector are not renormalized. As we saw, this does
not affect the cancellation of classical Newtonian singularity.

Another possibility is to look for some relation with the
(super)renormalizability of the theory. Indeed, that the
singularities are canceled in the renormalizable theory
fourth-derivative is known already from the first works of
Stelle \cite{Stelle-77,Stelle-78}.
We can say that this is also true for at least some of the
superrenormalizable models of \cite{highderi}, and for the
non-local model of \cite{Tomboulis,Modesto}. So, in reality
there is a strong correspondence between quantum and classical
properties in this case. However, the complete answer to
this question is possible only after further analysis of
the problem, taking into account the theories with complex
and multiple poles in the propagator. One observation
concerning this issue has been done recently by one of us
in \cite{CountGhosts}. The cancelation of Newtonian
singularity in the exponential gravity of \cite{Tseytlin-95}
and \cite{Tomboulis} can be actually seen as an effect of
an infinite amount of hidden ghost-like complex-poles states
in this theory,  presumably acting in a way similar to Eqs.
(\ref{poten}) and (\ref{posc}). This situation represents an
additional strong motivation to explore all aspects of the
theories with complex poles with a special attention.

%%%%%%%%%%%%%%%%%%%%%%%%%%%%%%%%%%%%%%%%%%%%%%%%%%%%%%%%%%%%%%%%%
\acknowledgments
The work was started during the visit of I.Sh. to Shanghai
and authors are very grateful to the Department of Physics
of Fudan University for organizing and supporting this visit.
T.P.N is grateful to Department of Physics of University of Alberta
and is especially grateful to Profs. A. Zenilkov and V.P. Frolov
for kind hospitality and support during his stay at the University
of Alberta. T.P.N. is grateful to CAPES
and Natural Sciences and Engineering Research Council of Canada
for supporting his visit to the University of Alberta.
%%%%%%%%%%%%%%%%%%%%%%%%%%%%%%%%%%%%%%%%
I.Sh. is very grateful to the D\'epartement de Physique Th\'eorique
and Center for Astroparticle Physics of Universit\'e de Gen\`eve
for support and kind hospitality during his sabbatical stay, and
to CNPq and also to FAPEMIG and ICTP for partial support of his work.
The authors are grateful to Breno Loureiro Giacchini for finding misprints in 
formulas in the end of Sect. 3, compared to the published version.

%%%%%%%%%%%%%%%%%%%%%%%%%%%%
%%  \paragraph{Note added.} This is also a good position for notes
%%  added after the paper has been written.

% The bibliography will probably be heavily edited during typesetting.
% We'll parse it and, using the arxiv number or the journal data, will
% query inspire, trying to verify the data (this will probalby spot
% eventual typos) and retrive the document DOI and eventual errata.
% We however suggest to always provide author, title and journal data:
% in short all the informations that clearly identify a document.

%%%%%%%%%%%%%%%%%%%%%%%%%%%%%%%%%%%%%%%%%%%%%%%%%%%%%%%%%%%%%

\end{document}